\documentclass[noshowpacs,nofootinbib,10pt]{revtex4}
\usepackage{mathtools}
\usepackage{color}
\usepackage{subcaption}
\usepackage{graphicx}
\usepackage{graphicx,float}\usepackage[all]{xy}
\usepackage{amsmath,upgreek,tikz,bm}
 
\usepackage{amssymb}

\usepackage{textgreek}

\newcommand{\beq}{\begin{eqnarray}}
\newcommand{\eeq}{\end{eqnarray}}
\newcommand{\be}{\begin{equation}}
\newcommand{\ee}{\end{equation}}

\newcommand{\bea}{\begin{eqnarray}}
\newcommand{\eea}{\end{eqnarray}}
\newcommand{\bes}{\begin{subequations}}
\newcommand{\ees}{\end{subequations}}

\newcommand{\ba}{\begin{eqnarray}}
\newcommand{\ea}{\end{eqnarray}}
\newcommand\orcidroldao{{\href{https://orcid.org/0000-0003-3978-532X}{\orcidicon}}}
\newcommand{\orcidicon}{%
	\begin{tikzpicture}
	\draw[lime, fill=lime] (0,0)
		circle [radius=0.16]
		node[white] {{\fontfamily{qag}\selectfont \tiny ID}};
	\draw[white, fill=white] (-0.0625,0.095)
		circle [radius=0.007];
	\end{tikzpicture}	\hspace{-2mm}
}

\usepackage[colorlinks,hyperindex,unicode]{hyperref}
\definecolor{green1}{RGB}{0,128,0} 
\hypersetup{hidelinks,pagebackref=true,hyperindex=true,colorlinks=true,breaklinks=true,urlcolor= blue}
\hypersetup{%
  colorlinks = true,
  linkcolor  = blue,
  citecolor = cyan,
}
\usepackage{bookmark,textgreek}
\usepackage{hyperref,color,xcolor}
\hypersetup{hidelinks,hyperindex=true,colorlinks=true,breaklinks=true,urlcolor= blue}
\hypersetup{%
  colorlinks = true,
  linkcolor  = blue
}

\begin{document}
\title{Echoes of the gravitational decoupling: scalar perturbations and quasinormal modes of  hairy black holes}

\author{R. T.  Cavalcanti}
\email{rogerio.cavalcanti@ufabc.edu.br} 
\affiliation{Center of Mathematics, Federal University of ABC,  09210-580, Santo Andr\'e, Brazil.\\ DFI, Universidade Estadual Paulista, Unesp, Guaratinguet\'{a}, 12516-410, Brazil}
\author{R. C. de Paiva}
\email{ronaldo.paiva@unesp.br} \affiliation{DFI, Universidade Estadual Paulista, Unesp, Guaratinguet\'{a}, 12516-410, Brazil}
\author{R. da Rocha\orcidroldao\!\!}
\email{roldao.rocha@ufabc.edu.br}
\affiliation{Center of Mathematics,  Federal University of ABC, 09210-580, Santo Andr\'e, Brazil.}



\begin{abstract} 
The behavior of hairy black hole solutions, obtained by the gravitational decoupling (GD) method, is investigated under scalar perturbations. The quasinormal mode frequencies of such solutions are regulated by GD hair. The numerically generated wave solutions are derived for a range of values for the GD hairy black hole parameters, with higher-frequency modes very sensitive to them. The results are confronted with the corresponding ones for the Schwarzschild solution, whose deviations from it demonstrate a unique physical identification of GD hairy black holes. The method here presented comprises the first steps towards the obtainment of the observable signature of GD hairy black holes at ground-based detectors, emitted from coalescing binary systems of GD hairy black hole mergers in the ringdown phase. 
\end{abstract}


\maketitle

\section{Introduction}

Gravitational decoupling (GD) methods comprise established successful protocols used to generate analytical solutions of Einstein's effective field equations \cite{Ovalle:2017fgl,Ovalle:2019qyi,Casadio:2012rf,Ovalle:2017wqi,Antoniadis:1998ig}, with plenty of relevant applications in gravitational physics and astrophysics, also including analog gravity models. 
 The gravitational decoupling and some extensions were studied in Refs. \cite{Casadio:2012rf,Ovalle:2014uwa,Ovalle:2016pwp,Ovalle:2013vna,Ovalle:2013xla,Ovalle:2018vmg,Casadio:2012pu,Casadio:2015jva}, and have been applied to construct new physical solutions from root solutions of general relativity (GR), that especially include an appropriate description of anisotropic stellar distributions \cite{daRocha:2020rda,Fernandes-Silva:2017nec,Ovalle:2007bn,Sharif:2018tiz,Morales:2018urp,Rincon:2019jal,Hensh:2019rtb,Ovalle:2019lbs,Gabbanelli:2019txr,Ovalle:2008se,Gabbanelli:2018bhs,Panotopoulos:2018law,Heras:2018cpz,Contreras:2018vph,Maurya:2020djz,Tello-Ortiz:2020euy}. Refs. \cite{daRocha:2020jdj,Fernandes-Silva:2019fez,daRocha:2017cxu,daRocha:2019pla} derived accurate physical constraints on the parameters in gravitational decoupled solutions, using the WMAP and also data at LIGO/Virgo \cite{Abbott:2016blz,Abbott:2017oio}.  
The gravitational decoupling procedure iteratively fabricates, upon a given isotropic source of gravitational field, anisotropic compact sources of gravity, that are weakly coupled. When starting by a perfect fluid, one can couple it to more elaborated types of stress-energy-momentum tensors, that underlie realistic compact configurations \cite{Contreras:2021yxe,Maurya:2019hds,Tello-Ortiz:2020svg,Maurya:2021fuy,Maurya:2021sju,Arias:2020hwz,Maurya:2019kzu,PerezGraterol:2018eut,Morales:2018nmq,Contreras:2019iwm,Arbanil:2021ahh,Maurya:2021aio,Singh:2019ktp,Muneer:2021lfz,Tello-Ortiz:2019gcl,Maurya:2019sfm,Sharif:2018toc,Estrada:2018zbh,Torres:2019mee,Abellan:2020jjl,Estrada:2018vrl,Leon:2019abq,Casadio:2019usg,Sharif:2019mjn,Abellan:2020wjw,Rincon:2020izv,Sharif:2020arn}. Particularly the case of GD involving hairy solutions has been explored, with astonishing and relevant physical ramifications \cite{Ovalle:2021jzf,Ovalle:2020kpd,Meert:2020sqv,Meert:2021khi}. 
Hairy black holes present additional degrees of freedom of macroscopic nature, which are not associated with a Gauss flux theorem for gravity. Hence these degrees of freedom are not associated with quasilocal
conserved quantities that can be computed at the GD hairy black hole event horizon level. The way how the 
microscopic description of black holes catches the additional degrees of freedom, corresponding to hair, points to an effective procedure for deriving analytical solutions. Therefore 
thermodynamical quantities of GD hairy black holes can be studied.
With the first unambiguous observations of gravitational waves at LIGO/Virgo, directly encoding smoking gun disturbances in the spacetime curvature, GD hairy solutions can be proposed and investigated also in the gravitational wave astrophysical scenario, exploring aspects of gravity in the strong nonlinear regime and comparing any deviation from the general-relativistic setup. 
Coalescing binary black hole systems in the merger ringdown phase, which have been scrutinized from the direct observational point of view with outstanding results \cite{LIGOScientific:2018dkp,LIGOScientific:2019fpa}, can shed new light on GD hairy solutions. 

In recent years, numerical relativity has been consolidated as an essential tool in the fully nonlinear regime of GR, supporting paramount results also arising from the post-Newtonian regime and black hole perturbation theory \cite{Barreto:2004fn}. Besides the huge progress in the field, since GR is not a complete theory \cite{Rovelli:1997yv, krasnov2020formulations,1984waldbook,Wald:1999vt}, persistent issues on its foundations do remain, such as the physical meaning of singularities and the lack of a QFT of gravity. These issues also motivated a considerable number of alternative theories focused on addressing them \cite{Vishwakarma:2016tzw}. The spectrum of the extensions of GR black holes, together with the improved sensibility of the future gravitational wave antennas, may constitute a set of unprecedented tests of foundations of gravity itself \cite{Carson2019,Carson2020, ni2005empirical,will2014confrontation}.  
Since the seminal work of Regge and Wheeler 
\cite{Regge_Wheeler_1957, chandrasekhar1998mathematical}, black hole perturbation theory has been playing a fundamental role in investigating problems in the strong field regime of GR, as well as in aspects of astrophysics, high-energy physics, and the foundations of the gravitational interaction as well \cite{Pani:2013,Cardoso:2019rvt,Barack:2018yly,Pani:2009ss}. With the growing interest in gravitational waves, black hole perturbation theory has been brought back to the spotlights. 
 The advances on black hole perturbation theory and its applications constitute now part of the essential tools of, but not restricted to, gravitational physics, from theory to observations \cite{Cardoso:2019rvt,Pani:2013,Barack:2018yly}. Among the main results, quasinormal modes quantify the black hole relaxation after any external perturbation \cite{chandrasekhar1998mathematical}. Perturbation theory also accounts for binary black hole mergers, which can emit radiation and, at the last
end-stage, the black hole orbits fall off, yielding increments in the
amplitude of the gravitational radiation. By emitting gravitational waves and the orbital period diminishing, black holes inspiral and merge into a stable end-stage, through the ringdown phenomenon. Before the advent of numerical relativity, perturbative methods were  the best one could get from realistic models of GR beyond the highly symmetric cases in the nonlinear regime. Quasinormal modes of black
holes can specify the stability of the background spacetime with respect to perturbations, yielding unique physical signatures.  
Quasinormal modes consist of wave modes of energy dissipation of scattering fields with different spin in a black hole background \cite{Leaver_Chandrasekhar_1985}, being formally derived from linearized differential equations of GR constraining the perturbations around a black hole solution. They are indeed governed by linear second-order partial differential
equations, complemented by suitable 
boundary conditions at the black hole event horizon and the spatial infinity as well. Quasinormal modes computed in the linearized setup present full compliance to those derived by a nonlinear coupled system of Einstein's equations, at least in what concerns sufficiently late times \cite{Richartz:2015saa,bishop2016extraction,KokkotasSchmidt:1999,Nollert:1999ji,BertiCardoso:2009,Konoplya:2011qq}. 

Quasinormal ringing dominates the majority of phenomena regarding black hole perturbation. Hence, quasinormal modes can 
impart exclusive imprints that yield the unequivocal observational identification of GD hairy black holes. To bring substantial information out of gravitational-wave detectors, one must thoroughly know the main features and behavior of quasinormal modes for GD hairy black holes. A prominent feature of this apparatus is the fact that the event horizon emulates a membrane for classical fields, yielding a non-Hermitian boundary value problem with complex eigenfrequencies. The imaginary part of the frequency encodes the decay timescale of the
black hole perturbation and quantifies the energy lost by the black hole. Perturbed black holes are inherently dissipative, due to event horizon effects, yielding quasinormal comprise universal phenomena in black hole physics. They predominate the radiation emission along the intermediate stages of the black hole perturbation. 
Those modes, also known as ringdown modes, are composed of a simple linear superposition of exponentially damped sinusoids, followed by a power-law tail \cite{Leaver_Chandrasekhar_1985, chandrasekhar1998mathematical}. A noticeable fact about quasinormal modes regards their specific dependence on black hole parameters, being thus an important tool for confronting theory and observations. In that sense, the spectrum of compact objects in extensions of GR can reveal possible observational signatures of alternative theories of gravity, as the GD setup \cite{Cardoso:2019rvt,Barack:2018yly}. Quasinormal frequencies are independent of the way that either the black hole itself or a given field surrounding it
are perturbed, being thoroughly and exclusively characterized by the black hole parameters. Therefore the main goal here is to derive the quasinormal spectrum of frequencies of a GD hairy black hole, analyzing the resulting deviations from the Schwarzschild standard solution when the GD hairy parameters are set in.

This paper is organized as follows: Section \ref{Sgd} is dedicated to reviewing the gravitational decoupling procedure, obtaining a metric for GD hairy black holes. In Section \ref{sPerturb} the scalar perturbations of GD hairy black hole solutions are investigated and discussed. It includes an analysis of the role played by the GD hair parameters, which regulate the resulting quasinormal modes. The waveform evolution of the mentioned solutions is also numerically found. Section \ref{4} is dedicated to conclusions and discussion.

\section{Gravitational decoupling algorithm and hairy black hole solutions}
\label{Sgd}
The GD algorithm can be implemented when one takes root solutions of Einstein's field equations to disentangle a given energy-momentum
tensor into subparts that are computationally more tractable \cite{Ovalle:2017wqi,Ovalle:2019qyi,Ovalle:2020kpd}. Einstein's field equations read 
\begin{equation}
\label{corr2}
G_{\mu\nu}
=
8\pi\,\check{T}_{\mu\nu},
\end{equation}
where $G_{\mu\nu}=
R_{\mu\nu}-\frac{1}{2}R g_{\mu\nu}$ denotes the Einstein tensor, and the energy-momentum tensor can be split off as 
\begin{equation}
\label{emt}
\check{T}_{\mu\nu}
=
{\scalebox{.91}{$\mathsf{T}$}}^{\rm}_{\mu\nu}
+
{\scalebox{.9}{${\alpha}$}}\,{\scalebox{.9}{${\Theta}$}}_{\mu\nu}.
\end{equation}
Eq. (\ref{emt}) regards a usual general-relativistic solution ${\scalebox{.91}{$\mathsf{T}$}}_{\mu\nu}$, whereas ${\scalebox{.9}{${\Theta}$}}_{\mu\nu}$ denotes ancillary sources residing in the gravitational sector, insomuch as ${\scalebox{.9}{${\alpha}$}}$ denotes the GD parameter. Also, the conservation equation $
\nabla_\mu\,\check{T}^{\mu\nu}=0$ holds, in general. 
One assumes an arbitrary static and spherically symmetric metric,  
\begin{equation}
ds^{2}
=
-e^{\gamma (r)}dt^{2}+e^{\upbeta (r)}dr^{2}
+r^{2}d\Upomega^2, 
\label{metric}
\end{equation}
which can be replaced into the system (\ref{corr2}), yielding 
\bes
\begin{eqnarray}
\label{ec1}
\!\!\!\!\!\!\!\!\!\!8\pi\!
\left(
{\scalebox{.91}{$\mathsf{T}$}}^0_{\ 0}+{\scalebox{.9}{${\alpha}$}}{\scalebox{.9}{${\Theta}$}}^0_{\ 0}
\right)
\!\!&\!=\!&\!
-\frac 1{r^2}
+
e^{-\upbeta }\left( \frac1{r^2}-\frac{\upbeta'}r\right),
\\
\label{ec2}
\!\!\!\!\!\!\!\!\!\!\!\!\!8\pi\!
\left({\scalebox{.91}{$\mathsf{T}$}}^1_{\ 1}+{\scalebox{.9}{${\alpha}$}}{\scalebox{.9}{${\Theta}$}}^1_{\ 1}\right)
\!\!&\!=\!&\!
-\frac 1{r^2}
+
e^{-\upbeta }\left( \frac 1{r^2}+\frac{\gamma'}r\right),
\\
\label{ec3}
\!\!\!\!\!\!\!\!\!\!\!\!\!\!\!\!\!\!\!\!\!\!\!\!\!\!\!\!\!\!\!\!\!\!\!\!\!8\pi\!
\left({\scalebox{.91}{$\mathsf{T}$}}^2_{\ 2}\!+\!{\scalebox{.9}{${\alpha}$}}{\scalebox{.91}{${\scalebox{.91}{${\Theta}$}}$}}^2_{\ 2}\right)
\!\!&\!=\!&\!
\!\frac {e^{-\upbeta }}{4}
\left(2\gamma''\!+\!\gamma'^2 \!-\upbeta'\gamma'
\!+\!\frac{1}{r}(\gamma' \!-\!\upbeta')\right)
\end{eqnarray}
\ees
for the prime denoting the derivative with regard to $r$.
The coupled system of ODEs (\ref{ec1}) -- (\ref{ec3}) are related to the density and the tangential/radial pressures, respectively by \cite{Ovalle:2017wqi,Ovalle:2019qyi}
\bes
\beq
\check{\rho}
&=&
\rho+
{\scalebox{.9}{${\alpha}$}}{\scalebox{.9}{${\Theta}$}}^0_{\ 0},\label{efecden}\\
\check{p}_{t}
&=&
p
-{\scalebox{.9}{${\alpha}$}}{\scalebox{.9}{${\Theta}$}}^2_{\ 2}, 
\label{efecpretan}\\\check{p}_{r}
&=&
p
-{\scalebox{.9}{${\alpha}$}}{\scalebox{.9}{${\Theta}$}}^1_{\ 1}.
\label{efecprera}
\eeq\ees With these quantities defined, the system anisotropy is quantified by the parameter
\beq
\Updelta = 
\check{p}_{r}-\check{p}_{t}.\eeq 
Solving the coupled system \eqref{corr2} exclusively for the root source ${\scalebox{.91}{$\mathsf{T}$}}_{\mu\nu}$ is feasible when one considers solutions of type \cite{Ovalle:2017wqi,Ovalle:2020kpd} 
\begin{equation}
ds^{2}
=
-e^{\kappa (r)}dt^{2}
+e^{\upzeta (r)}dr^{2}
+
r^{2}d\Upomega^2
,
\label{pfmetric}
\end{equation}
where 
\begin{equation}
\label{standardGR}
e^{-\upzeta(r)}
\equiv
1-\frac{8\pi}{r}\int_0^r \mathsf{r}^2\,{\scalebox{.91}{$\mathsf{T}$}}^0_{\ 0}(\mathsf{r})\, d\mathsf{r}
=
1-\frac{2m(r)}{r}
\end{equation}
regards the Misner--Sharp mass function, which is the quasilocal mass within a sphere
of radius
$r$, at a given proper time, providing a quasilocal prescription of the energy that fabricates curvature accommodated into a black hole.
The auxiliary source ${\scalebox{.9}{${\Theta}$}}_{\mu\nu}$ induces the gravitational decoupling of the root metric~\eqref{pfmetric}, carried through the functions  
\bes
\begin{eqnarray}
\label{gd1}
\kappa(r)
&\mapsto &
\gamma(r)=\kappa(r)+{\scalebox{.9}{${\alpha}$}} f_2(r)
\\
\label{gd2}
e^{-\upzeta(r)} 
&\mapsto &
e^{-\upbeta(r)}=e^{-\upzeta(r)}+{\scalebox{.9}{${\alpha}$}} f_1(r)
, 
\end{eqnarray}
\ees
where $f_1(r)$ [$f_2(r)$] denotes the GD for the temporal [radial]  metric
component.
Eqs.~(\ref{gd1}, \ref{gd2}) split the Einstein's field equations~(\ref{ec1}) -- (\ref{ec3}) into two distinct arrays. 
The first one encodes the GR coupled system ${\scalebox{.91}{$\mathsf{T}$}}_{\mu\nu}$, solved by the root metric~(\ref{pfmetric}). The second one is associated to the GD tensor ${\scalebox{.9}{${\Theta}$}}_{\mu\nu}$ and reads
\bes
\begin{eqnarray}
\label{ec1d}
\!\!\!\!\!8\pi\,{\scalebox{.9}{${\Theta}$}}^0_{\ 0}
&=&
{\scalebox{.9}{${\alpha}$}}\left(\frac{f_1}{r^2}+\frac{f_1^\prime}{r}\right),
\\
\label{ec2d}
\!\!\!\!\!8\pi\,{\scalebox{.9}{${\Theta}$}}^1_{\ 1}
-{\scalebox{.9}{${\alpha}$}}\,\frac{e^{-\upzeta}\,f_2^\prime}{r}
&\!=\!&
{\scalebox{.9}{${\alpha}$}}\,f_1\left(\frac{1}{r^2}+\frac{\gamma'}{r}\right)
\\
\label{ec3d}
\!\!\!\!\!\!\!\!\!\!\!\!\!8\pi{\scalebox{.9}{${\Theta}$}}^2_{\ 2}\!-\!{\scalebox{.9}{${\alpha}$}}{f_1}\left(2\gamma''\!+\!\gamma'^2\!+\!\frac{2\gamma'}{r}\right)\!&\!\!=\!\!&\!\!\!{\scalebox{.9}{${\alpha}$}}\frac{f_1^\prime}{4}\!\left(\gamma'\!+\!\frac{2}{r}\right)\!+\!V
\end{eqnarray}
\ees
where \cite{Ovalle:2017fgl}
\beq
V(r) = {\scalebox{.9}{${\alpha}$}} e^{-\upzeta}\left(2f_2^{\prime\prime}+f_2^{2\prime}+\frac{2f_2^\prime}{r}+2\kappa' f_2^\prime-\upzeta' f_2^\prime\right).\eeq
The tensor-vacuum, defined when the conditions ${\scalebox{.9}{${\Theta}$}}_{\mu\nu}\neq 0$ and ${\scalebox{.91}{$\mathsf{T}$}}_{\mu\nu}=0$ simultaneously hold, leads to GD hairy black hole solutions \cite{Ovalle:2018umz,Ovalle:2022yjl}. 
Eqs. (\ref{ec1}) -- (\ref{ec3}) then yield the radial pressure to attain negative values, 
\begin{equation}
\check{p}_{r}
=
-\check{\rho}.
\label{schwcon}
\end{equation}
and, together to the Schwarzschild solution, it implies that 
\begin{equation}
\label{fg}
{\scalebox{.9}{${\alpha}$}}\,f_1(r)
=
\left(1-\frac{2M}{r}\right)\left(e^{{\scalebox{.65}{${\alpha}$}}\,f_2(r)}-1\right)
,
\end{equation}
so that \eqref{metric} becomes
\begin{eqnarray}
\label{hairyBH}
\!\!\!\!ds^{2}
\!\!&\!=\!&\!\!
-\left(1-\frac{2M}{r}\right)
e^{{\scalebox{.65}{${\alpha}$}} f_2(r)}
dt^{2}
\!+\!\left(1-\frac{2M}{r}\right)^{-1}\!\!\!\!
e^{-{\scalebox{.65}{${\alpha}$}}\,f_2(r)}
dr^2
\nonumber
\\
&&
\qquad\qquad\qquad\qquad\qquad\qquad\qquad+r^{2}\,d\Upomega^2
.
\end{eqnarray}
In the radial range $r\geq 2M$, the tensor-vacuum is given by expressing ${\scalebox{.9}{${\Theta}$}}^0_{\ 0}$ by the most general linear combination of the radial and tangential components of the stress-energy-momentum tensor, as 
\beq\label{comb}
{\scalebox{.9}{${\Theta}$}}^0_{\ 0}
=
a\,{\scalebox{.9}{${\Theta}$}}^1_{\ 1}+b\,{\scalebox{.9}{${\Theta}$}}^2_{\ 2}.\eeq 
Eqs.~(\ref{ec1d}) -- (\ref{ec3d}) then yield  
\begin{eqnarray}
\label{master}
\!\!\!\!\!\!\!\!\!\!\!\!\!\!\!\!\!\!&&r(r-2M)bh''+2\left[(b+a-1)r+2(1-a)M\right]
h'
\nonumber
\\
\!\!\!\!\!\!\!\!\!\!\!\!\!\!\!&&
\qquad\qquad+2\,(a-1)(h-1)=0,
\end{eqnarray}
for 
\beq
h(r)
=
e^{{\scalebox{.65}{${\alpha}$}}\,f_2(r)}.\eeq A trivial deformation corresponding to the standard Schwarzschild solution can be yielded when $a = 1$.
The solution of Eq. (\ref{master}) can be written as \cite{Ovalle:2017fgl}
\begin{equation}
\label{master2}
e^{{\scalebox{.65}{${\alpha}$}}\,f_2(r)}
=
1+\frac{1}{r-2M}
\left[{{\mathfrak{l}}}_0+r\left(\frac{{{\mathfrak{l}}}}{r}\right)^{m}
\right]
,
\end{equation}
where ${{\mathfrak{l}}}_0={\scalebox{.9}{${\alpha}$}}{{\mathfrak{l}}}$ is a primary hair charge, whereas \beq\label{nnn}
m
=
\frac{2}{b}\left(a-1\right),\eeq
with $m>1$ for the asymptotic flatness condition. 
On the other hand, the dominant energy conditions, \beq
\check{\rho}\geq |\check{p}_t|,\quad \check{\rho}\geq|\check{p}_r|,\label{dec0}\eeq yield $m\le2$, therefore restricting the parameter $a$ and $b$ by the equations (\ref{comb}) and (\ref{nnn}) \cite{Ovalle:2017wqi,Ovalle:2019qyi,Ovalle:2020kpd}.
 Besides, the strong energy conditions, 
 \bes
\begin{eqnarray}
\check{\rho}+2\,\check{p}_t+\check{p}_r
\geq
0, 
\label{strong01}\\
\check{p}_r+\check{\rho}
\geq
0,\\
\check{p}_t+\check{\rho}
\geq
0,
\end{eqnarray}
\ees
make Eq.~\eqref{schwcon} to read 
\beq\label{t001}
-\,{\scalebox{.9}{${\Theta}$}}^0_{\ 0}\leq{\scalebox{.9}{${\Theta}$}}^2_{\ 2}\leq0.\eeq Therefore, together with Eqs.~\eqref{ec1d} and~\eqref{ec3d}, Eq. (\ref{t001}) can be recast as \bes\begin{eqnarray}
\label{strong5}
\!\!\!\!\!\!\!\!\!\!\!\!\!\!\!\!\!\!G_1(r)&:=&{h''(r-2M)+2h'}
\geq
0,\\
\label{strong51}
\!\!\!\!\!\!\!\!\!\!\!\!\!\!\!\!\!\!G_2(r)&:=&
h''(r-2M)r
+4h'M
-2h+2
\geq
0.
\end{eqnarray} \ees
The mapping 
\begin{equation}
\label{gauge}
h(r)
\mapsto
h(r)-\frac{{{\mathfrak{l}}}_0}{r-2M}
\end{equation}
leaves $G_1(r)$ and $G_2(r)$ invariant. 
Solutions with a proper horizon at $r\sim 2M$ and 
approach the standard Schwarzschild solution in the $r\gg 2M$ regime, implying that $
G_1(r)=0$. Hence, solving Eq.~\eqref{strong5} implies that  
\begin{equation}
\label{strongg}
h(r)
=
d
-
{\scalebox{.9}{${\alpha}$}}\,\frac{{{\mathfrak{l}}}-r\,e^{-r/M}}{r-2M},
\end{equation} where $d$ is an arbitrary integration constant. 
Also, Eq.~\eqref{strongg} depends upon the inequality \eqref{strong51}. Replacing~\eqref{strongg} into \eqref{hairyBH} implies the metric 
\begin{equation}
\label{strongBH}
e^{\gamma}
=
e^{-\upbeta}
=
1-\frac{2\mathsf{M}}{r}+{\scalebox{.9}{${\alpha}$}}\,e^{-r(\mathsf{M}-{\scalebox{.65}{${\alpha}$}}\,{{\mathfrak{l}}}/2)^{-1}}
,
\end{equation}
to represent a hairy black hole, 
where $\mathsf{M}=M+{\scalebox{.9}{${\alpha}$}}\,{{\mathfrak{l}}}/2$.

Now, the strong energy conditions are consistent with ${{\mathfrak{l}}}
\geq
2M/e^{2}$, whose extremal case ${{\mathfrak{l}}}=2M/e^{2}$ leads to 
the GD hairy black hole metric, \begin{eqnarray}
\label{strongh2M}
e^{\gamma}
=
e^{-\upbeta}
=
1-\frac{2M}{r}+{\scalebox{.9}{${\alpha}$}}\left(e^{-r/M}-\frac{2M}{e^2\,r}\right),
\end{eqnarray}
which has the horizon at $r_{\scalebox{.56}{\textsc{hor}}} = 2M$. 
In what follows the quasinormal modes of the GD hairy black hole \eqref{strongh2M} will be derived and discussed.

\section{Scalar perturbations}{\label{sPerturb}}

\begin{figure}[H]
 \centering
\centering
	\includegraphics[width=.65\textwidth]{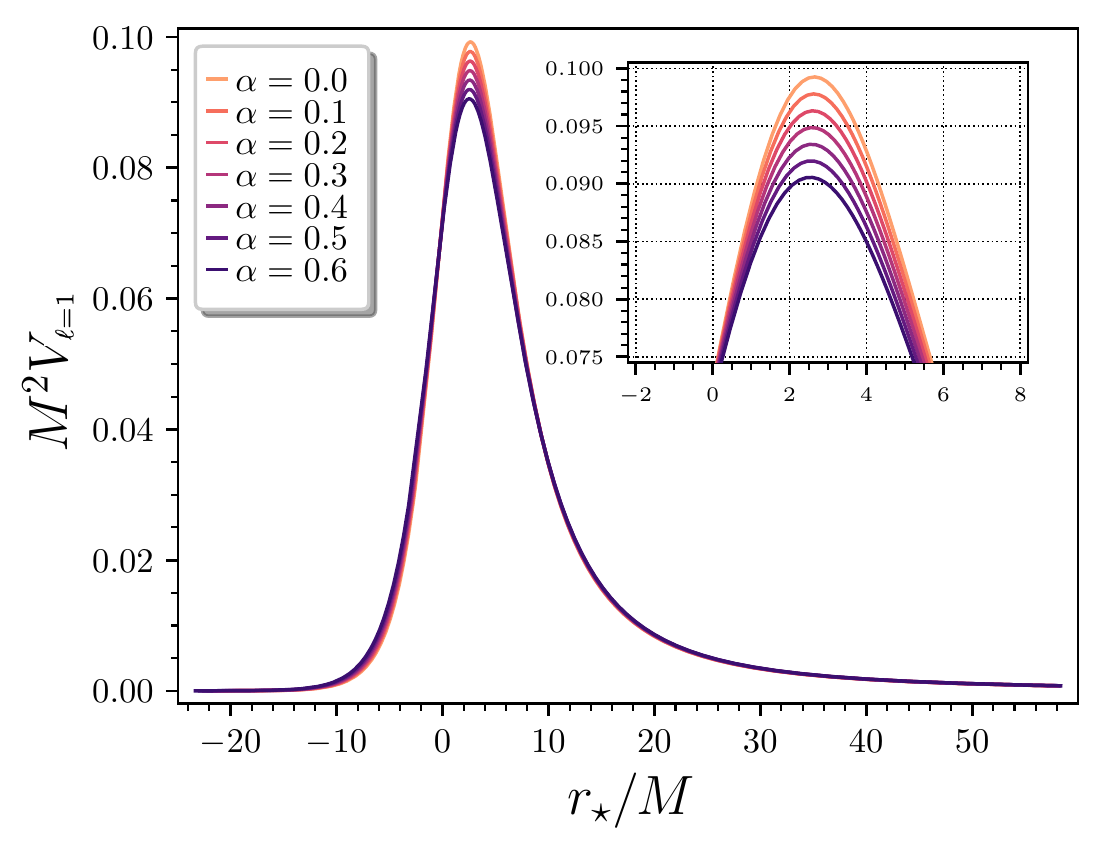}
 \caption{Effective potential, in tortoise coordinate, for $\ell\!=\!1$, in the extreme case $({{\mathfrak{l}}} = 2M/e^2)$, for varying ${{\scalebox{.9}{${\alpha}$}}}$. The ${{\scalebox{.9}{${\alpha}$}}}=0$ corresponds to the Schwarzschild case.}
 \label{extremePotential}
\end{figure}

External perturbations can excite the quasinormal modes of GD hairy black holes, appearing as damped oscillations in the black hole natural response. These vibrations are inherent features of the exterior geometry (\ref{strongh2M}), uniquely identifying GD hairy black holes. 
The aim in this section is to compare, both qualitatively and quantitatively, the spectrum of scalar perturbations of the metric given in (\ref{strongBH}) and (\ref{strongh2M}) to the well-known results of the Schwarzschild black hole, looking for possible preliminary signatures of GD hairy black holes. 
We start by considering the GD hairy black hole metric (\ref{strongBH})
and a minimally coupled massless scalar field $\Phi$ propagating in such background, whose dynamics is described by the Klein--Gordon equation,
\begin{align}\label{kg}
 \nabla^\mu\nabla_\mu \Phi = 0.
\end{align}
Solutions of (\ref{kg}) are composed by modes of frequency $\omega$, orbital number $\ell \in \mathbb{N}$, and azimuthal $m \in \mathbb{Z}$, with $-\ell \leq m \leq \ell$, with ansatz
\begin{align}\label{phiAnsatz}
 \Phi_{\omega \ell m}(t, r, \theta, \phi) = \exp({i\omega t - im\phi}) \frac{\Psi_{\omega \ell}(r)}{r}S_{\ell m}(\theta),
\end{align}
where $S_{\ell m}(\theta)$ denotes the angular part, in which we are not interested for now. 
As usual, generalized tortoise coordinates are used, 
\begin{align}\label{tortoise}
 \dfrac{dr_\star}{dr} = \left({1 - \dfrac{{{\scalebox{.9}{${\alpha}$}}} {{\mathfrak{l}}} + 2M}{r} + {{\scalebox{.9}{${\alpha}$}}} e^{-\frac{r}{M}}}\right)^{-1}.
\end{align}
The resulting equation takes the form of a Schr\"odinger-like equation,
\begin{align}\label{schrodinger-like equation}
\Psi''_{\omega \ell}\left(r_\star\right) +\left[\omega^2 - V_{\ell}\left(r,{{\scalebox{.9}{${\alpha}$}}},{{\mathfrak{l}}}\right)\right] \Psi_{\omega \ell}\left(r_\star\right) = 0,
\end{align}
with effective potential \cite{Cardona:2017scd,Beyer:1998nu}, in the case of GD hairy black holes, given by 
\begin{widetext}
\beq\label{effPot1}
 V_\ell(r,{{\scalebox{.9}{${\alpha}$}}}, \mathfrak{l}) &=& 
 {\left(1-\frac{2M}{r}\right)}{\left(\frac{{\left(\ell + 1\right)} \ell}{r^2} 
 + \frac{2M}{r^3}\right)}
 +\frac{{\scalebox{.9}{${\alpha}$}}}{r}\left[\frac{\left(\ell(\ell+1) + 2\right)}{r}e^{-\frac{r}{M}}
 - \frac{e^{-\frac{r}{M}}}{M} 
 - \frac{4M\mathfrak{l}}{r^3} 
 - \frac{\mathfrak{l}}{r^{2}}\left(\ell(\ell+1)
 - 1 - e^{\frac{2M - r}{M}} \right)\right]\nonumber\\
&& \qquad\qquad\qquad\qquad\qquad\qquad\quad\;\;+ \frac{{{\scalebox{.9}{${\alpha}$}}}^2}{r} \left[
 \frac{\mathfrak{l}}{M r^2}(M+r)e^{-\frac{r}{M}} 
 - \frac{e^{-\frac{2r}{M}}}{M} 
 - \frac{{\mathfrak{l}}^{2}}{r^{3}}\right].
\eeq
\end{widetext}
Notice that, similarly to the GD field equations, the potential can be naturally expressed as a polynomial in the parameter ${{\scalebox{.9}{${\alpha}$}}}$. The 0${}^{\rm th}$-order term is just the Schwarzschild potential and the corrections depend on the GD parameter ${{\mathfrak{l}}}$. The extreme case $({{\mathfrak{l}}} = 2M/e^2)$ will be also considered, with metric given by Eq. (\ref{strongh2M}) and effective potential reading 
\begin{widetext}
\begin{align}\label{effPot2}\nonumber
 V_\ell(r,{{\scalebox{.9}{${\alpha}$}}}) =& {\left(1-\frac{2M}{r}\right)}{\left(\frac{{\left(\ell + 1\right)} \ell}{r^{2}} + \frac{2M}{r^{3}}\right)} +\frac{{{\scalebox{.9}{${\alpha}$}}}}{r}{\left[\frac{\left(\ell(\ell+1) + 2\right)}{r}e^{-\frac{r}{M}} - \frac{e^{-\frac{r}{M}}}{M} - \frac{8M^{2} }{e^{2}r^{3}} - \frac{2M}{e^{2}{r^{2}}}{\left(\ell(\ell+1) - 1 - e^{\frac{2M - r}{M}}\right)}\right]} \\
& \qquad\qquad\qquad\qquad\qquad\qquad\quad\;+ \frac{{{\scalebox{.9}{${\alpha}$}}}^{2}}{r}{\left( \frac{2 M e^{-\frac{r + 2M}{M}}}{r^{2}} + \frac{2e^{-\frac{r + 2M}{M}}}{r} - \frac{e^{-\frac{2r}{M}}}{M} -\frac{4M^{2}}{e^{4}r^{3}}\right)}.
\end{align}
\end{widetext}
Fig. \ref{extremePotential} illustrates the effective potential \eqref{effPot2}, for several values of the GD parameter {{\scalebox{.9}{${\alpha}$}}}. 
The higher the value of the GD parameter ${{\scalebox{.9}{${\alpha}$}}}$, the lower the peak of the effective potential is. The GD attenuates the influence of the effective potential. 
For ${{\scalebox{.9}{${\alpha}$}}}=0.6$, the peak of the effective potential is 10\% lower than for the Schwarzschild case. Also,
 we numerically verified that the peak of the effective potential scales with ${{\scalebox{.9}{${\alpha}$}}}$, for $\ell=1$, as 
\beq
\!\!\!\!\!\!\!\!\!M^2V_{\scalebox{.52}{\textsc{peak}}}({{\scalebox{.9}{${\alpha}$}}})&=&0.02222\, {{\scalebox{.9}{${\alpha}$}}}^3-0.01726\,{{\scalebox{.9}{${\alpha}$}}}^2-0.01045\,{\scalebox{.9}{${\alpha}$}} + 0.09865,\nonumber
\eeq
within $0.001\%$ numerical error. One must additionaly emphasize that the higher the GD parameter ${{\scalebox{.9}{${\alpha}$}}}$, the smoother the profile of the effective potential is.

\subsection{The Mashhoon procedure and the Pöschl--Teller potential}
As the exact solution of the quasinormal modes is not easy to obtain analytically, approximation methods can be applied to derive them \cite{qian2021asymptotical}. Due to simplicity and reasonable results, the so-called Mashhoon method will be considered in what follows \cite{blome1984quasi,ferrari1984new}. Most of the intricacies to compute quasinormal modes of black hole solutions come from the potential that has a slow decay when asymptotically approaching the radial infinity. Owing to the existence of a branch cut, there is a backscattering phenomenon of gravitational waves off the potential, yielding late-time
tails. The Mashhoon procedure circumvent these complications and exact solutions can be derived when regarding a Pöschl--Teller that exponentially decays in the limit $r\to\infty$, also carrying fundamental features of the Schwarzschild effective potential.
The boundary conditions to solve the wave equation yield an evanescent  
scalar wave function at the boundary. It implies that computing the quasinormal modes can be qualified as calculating bound states for an inverse potential, $V\mapsto-V$, having a gap that can be suitably approximated by an effective Pöschl--Teller potential \cite{poschl1933bemerkungen}, 
\begin{equation}
V_{{\scalebox{.55}{\textsc{PT}}}}(r) = \frac{V_0}{\cosh^2
[\beta(r - r_0)]},
\end{equation}
with the quasinormal mode frequencies given by (see \cite{blome1984quasi,ferrari1984new,Konoplya:2011qq,BertiCardoso:2009} for details)
\begin{align}
 \omega = \omega_0 + i\Gamma,
\end{align}
where
\begin{align}
 \omega_0 = \sqrt{V_0 - \frac{\beta^2}{4}}, \qquad \Gamma = \beta\left(n + \frac{1}{2}\right),
\end{align}
where $n\in\mathbb{N}$ is the overtone number \cite{Konoplya:2011qq,BertiCardoso:2009}, and
\begin{align}
\beta^2 = -\frac1{2V_0}\dfrac{d^2V}{dr_\star^2}\bigg|_{r_0}.
\end{align}
%
Here $V_0$ denotes the effective potential maximum and $V_0 = V(r_0)$.

The Mashhoon method provides accurate results for some well-known black hole solutions, analytically computing quasinormal modes using their connection with bound
states of the black hole effective curvature potentials. Specifically, for gravitational perturbations of the 4-dimensional Schwarzschild black hole, the errors when calculating the fundamental quasinormal modes by this method do not exceed 2\% \cite{Konoplya:2011qq}. Table \ref{tabela} displays an array of quasinormal mode frequencies, for several values of $\ell$ and $n$, in the extreme potential case. For an arbitrary fixed value of the GD parameter ${{\scalebox{.9}{${\alpha}$}}}$, the higher the value of $\ell$, the bigger the real part of the eigenfrequency is. For fixed values of $\ell$, and consequently for fixed values of the real part of the frequencies of the quasinormal modes, the higher the overtone, the bigger the imaginary part of the eigenfrequency is. Now, for concomitantly fixed values of both $\ell$, the higher the value of the GD parameter ${{\scalebox{.9}{${\alpha}$}}}$, the lower both the real and imaginary parts of the frequencies of the quasinormal modes are.  

\begin{figure}[H]
 \centering
 \begin{center}
\includegraphics[width=.55\textwidth]{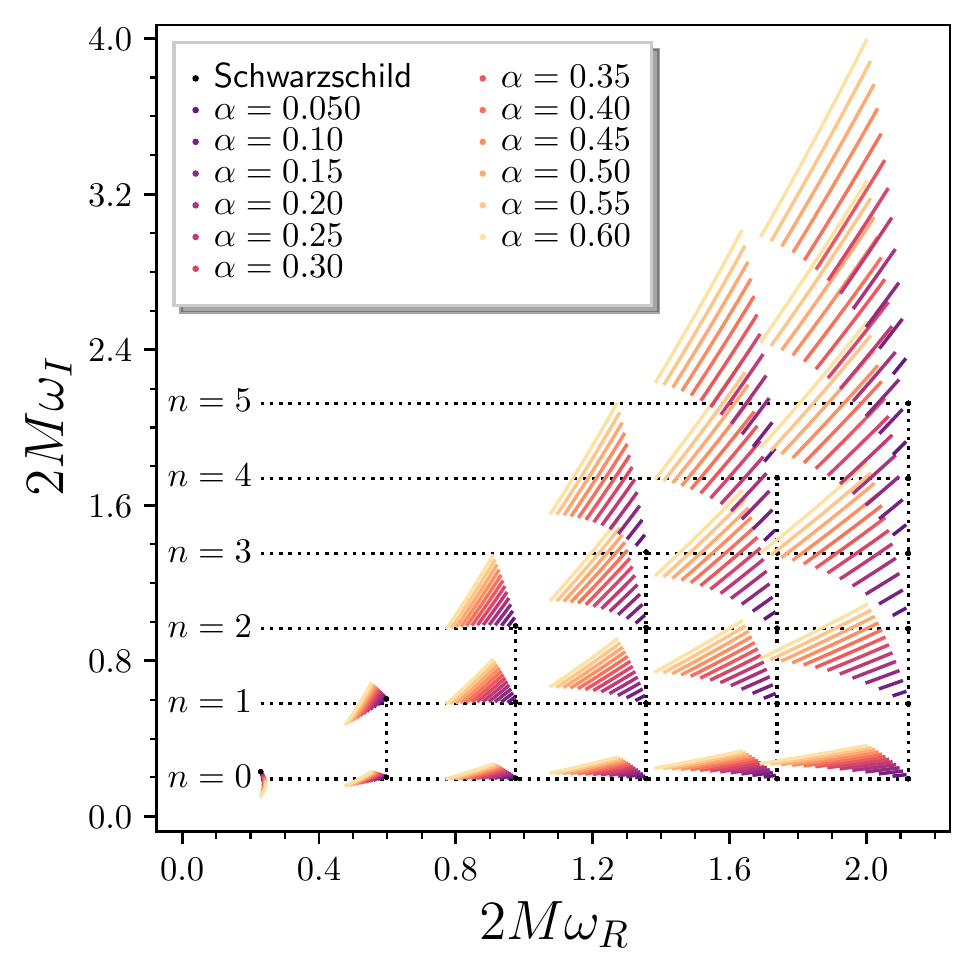}
\end{center}
 \caption{First quasinormal modes for the effective potential in the non-extreme regime. The lines for a given pair $(\ell,n)$ corresponds to a fixed ${{\scalebox{.9}{${\alpha}$}}}$ and varying $0.3\leq {\mathfrak{l}}\leq 0.9$. The vertical lines correspond, from left to right, to $\ell=1,2,3,4,5$ and the first point from left corresponds to $\ell=0$. }
 \label{qmnNonextreme}
\end{figure}

\begin{table}[H]
 \centering
 \begin{tabular}{c|c||c|c|c|c|c}
 \hline\hline
 ${\ell}$ & $\bf \emph{n}$ & $\bf {{\scalebox{.9}{${\alpha}$}}}=0.0$ & $\bf {{\scalebox{.9}{${\alpha}$}}}=0.1$ & $\bf {{\scalebox{.9}{${\alpha}$}}}=0.2$ & $\bf {{\scalebox{.9}{${\alpha}$}}}=0.3$ & $\bf {{\scalebox{.9}{${\alpha}$}}}=0.4$ \\ \hline\hline
 \,\,0 \,&\, 0 \,&\, 0.2296 + 0.2296$i$ \,&\, 0.2284 + 0.2254$i$ \,&\, 0.2271 + 0.2212$i$ \,&\, 0.2258 + 0.2169$i$ \,&\, 0.2245 + 0.2126$i$ \,\\ \hline
 \,\,1 \,&\, 0 \,&\, 0.5971 + 0.2013$i$ \,&\, 0.5933 + 0.1980$i$ \,&\, 0.5894 + 0.1947$i$ \,&\, 0.5855 + 0.1914$i$ \,&\, 0.5816 + 0.1881$i$\, \\ \hline
 \,\,1 \,&\, 1 \,&\, 0.5971 + 0.6039$i$ \,&\, 0.5933 + 0.5940$i$ \,&\, 0.5894 + 0.5841$i$ \,&\, 0.5855 + 0.5742$i$ \,&\, 0.5816 + 0.5643$i$\, \\ \hline
 \, 2 \,&\, 0 \,&\, 0.9747 + 0.1958$i$ \,&\, 0.9687 + 0.1928$i$ \,&\, 0.9625 + 0.1898$i$ \,&\, 0.9564 + 0.1867$i$ \,&\, 0.9502 + 0.1837$i$\, \\ \hline
 \, 2 \,&\, 1 \,&\, 0.9747 + 0.5874$i$ \,&\, 0.9687 + 0.5784$i$ \,&\, 0.9625 + 0.5693$i$ \,&\, 0.9564 + 0.5602$i$ \,&\, 0.9502 + 0.5512$i$\, \\ \hline
 \, 2 \,&\, 2 \,&\, 0.9747 + 0.9790$i$ \,&\, 0.9687 + 0.9639$i$ \,&\, 0.9625 + 0.9488$i$ \,&\, 0.9564 + 0.9337$i$ \,&\, 0.9502 + 0.9187$i$\, \\ \hline
 \, 3 \,&\, 0 \,&\, 1.3560 + 0.1942$i$ \,&\, 1.3480 + 0.1912$i$ \,&\, 1.3390 + 0.1883$i$ \,&\, 1.3311 + 0.1854$i$ \,&\, 1.3220 + 0.1824$i$\, \\ \hline
 \, 3 \,&\, 1 \,&\, 1.3560 + 0.5825$i$ \,&\, 1.3480 + 0.5737$i$ \,&\, 1.3390 + 0.5649$i$ \,&\, 1.3311 + 0.5561$i$ \,&\, 1.3220 + 0.5473$i$\, \\ \hline
 \, 3 \,&\, 2 \,&\, 1.3560 + 0.9709$i$ \,&\, 1.3480 + 0.9562$i$ \,&\, 1.3390 + 0.9415$i$ \,&\, 1.3311 + 0.9269$i$ \,&\, 1.3220 + 0.9122$i$\, \\ \hline
 \, 3 \,&\, 3 \,&\, 1.3560 + 1.3590$i$ \,&\, 1.3480 + 1.3390$i$ \,&\, 1.3390 + 1.3181$i$ \,&\, 1.3311 + 1.2980$i$ \,&\, 1.3220 + 1.2770$i$ \,\\ \hline\hline
 \end{tabular}
 \caption{Quasinormal modes frequencies $(2M\omega)$ for the extreme potential case computed using the Mashhoon procedure. The parameter ${{\scalebox{.9}{${\alpha}$}}}=0$ corresponds to the Schwarzschild case.}
 \label{tabela}
\end{table}
Fig. \ref{qmnNonextreme} depicts the quasinormal modes of GD hairy black holes for the effective potential in the non-extreme regime, with a large deviation  and the superposition of modes for increasing overtones $n$, in what concerns the Schwarzschild quasinormal mode frequencies. 

\subsection{The WKB approximation}

An alternative method for deriving quasinormal modes frequencies is to apply the well known WKB approximation \cite{BertiCardoso:2009,Santos:2019yzk}. In this section we use the third-order WKB approximation to derive the quasinormal frequencies of a GD hairy black hole. The quasinormal frequencies are given by \cite{Iyer:1986np, Iyer:1986nq,Santos:2019yzk}

\begin{align}
\omega^2 =& \left[V + \sqrt{(-2V^{(2)}}\Gamma\right] - i\sqrt{\delta(-2V^{(2)})}(1 +\Omega),
\end{align}
where 
\begin{align}
 \Gamma  =& \frac{1}{\sqrt{-2V^{(2)}}} \left[\frac{1}{8} \left(\frac{V^{(4)}}{V^{(2)}} \right)\left(\frac{1}{4} + \delta \right)\right.
 \left.- \frac{1}{288}\left(\frac{V^{(3)}}{V^{(2)}}\right)^2 (7 +60\delta) \right],
\end{align}

\begin{widetext}
\beq
 \Omega &= &-\frac{1}{2V^{(2)}} \left[\frac{5}{6912} \left(\frac{V^{(3)}}{V^{(2)}} \right)^2 (77 + 188\delta) - \frac{1}{384} \left(\frac{\left(V^{(3)}\right)^2 V^{(4)}}{\left(V^{(2)}\right)^3} (51 + 100\delta) \right)+\frac{1}{2304}\left(\frac{V^{(4)}}{V^{(2)}} \right)^2(67+68\delta)\right. \nonumber\\
&&\left.+\frac{1}{288}\left(\frac{V^{(3)} V^{(5)}}{\left(V^{(2)}\right)^2} \right)(19 + 28\delta)- \frac{1}{288}\left(\frac{V^{(6)}}{V^{(2)}} \right)(5+4\delta)\right],
 \eeq
\end{widetext}
and
\begin{align}
\delta = \left(\frac{1}{2} + n\right)^2.
\end{align}

Here $V^{(m)}$ denotes the $m-$th derivatives of the effective potential,
\begin{align}
V^{(m)} = \dfrac{d^m V}{dr_\star^m},
\end{align}
evaluated at potential's maximum. Table \ref{tabela2} displays the quasinormal frequencies, for several values of $\ell$ and $n$, in the extreme case. As expected, for fixed values of the GD parameter $\alpha$ and mode $\ell$, increasing the overtone $n$ increases the imaginary part of the eigenfrequency. Furthermore, as the imaginary part is associated with the damping rate, increasing the value of the GD parameter $\alpha$ makes the black hole  better behaves as an oscillator.

 
\begin{table}[H]
\centering
 \begin{tabular}{c|c||c|c|c|c|c}
 \hline\hline
 ${\ell}$ & $\bf \emph{n}$ & $\bf {{\scalebox{.9}{${\alpha}$}}}=0.0$ & $\bf {{\scalebox{.9}{${\alpha}$}}}=0.1$ & $\bf {{\scalebox{.9}{${\alpha}$}}}=0.2$ & $\bf {{\scalebox{.9}{${\alpha}$}}}=0.3$ & $\bf {{\scalebox{.9}{${\alpha}$}}}=0.4$ \\ \hline\hline
 \,\,0 \,&\, 0 \,&\, 0.2093 + 0.2304$i$ \,&\, 0.2071 + 0.2251$i$ \,&\, 0.2049 + 0.2197$i$ \,&\, 0.2028 + 0.2143$i$ \,&\, 0.2007 + 0.2089$i$ \,\\ \hline
 \,\,1 \,&\, 0 \,&\, 0.5822 + 0.1960$i$ \,&\, 0.5791 + 0.1930$i$ \,&\, 0.5759 + 0.1899$i$ \,&\, 0.5727 + 0.1869$i$ \,&\, 0.5695 + 0.1839$i$\, \\ \hline
 \,\,1 \,&\, 1 \,&\, 0.5244 + 0.6149$i$ \,&\, 0.5229 + 0.6041$i$ \,&\, 0.5214 + 0.5934$i$ \,&\, 0.5199 + 0.5828$i$ \,&\, 0.5186 + 0.5722$i$\, \\ \hline
 \, 2 \,&\, 0 \,&\, 0.9664 + 0.1936$i$ \,&\, 0.9607 + 0.1907$i$ \,&\, 0.9549 + 0.1878$i$ \,&\, 0.9491 + 0.1849$i$ \,&\, 0.9433 + 0.1820$i$\, \\ \hline
 \, 2 \,&\, 1 \,&\, 0.9264 + 0.5916$i$ \,&\, 0.9220 + 0.5823$i$ \,&\, 0.9176 + 0.5730$i$ \,&\, 0.9132 + 0.5637$i$ \,&\, 0.9088 + 0.5545$i$\, \\ \hline
 \, 2 \,&\, 2 \,&\, 0.8633 + 1.007$i$ \,&\, 0.8607 + 0.9903$i$ \,&\, 0.8580 + 0.9738$i$ \,&\, 0.8555 + 0.9574$i$ \,&\, 0.8528 + 0.9409$i$\, \\ \hline
 \, 3 \,&\, 0 \,&\, 1.350 + 0.1930$i$ \,&\, 1.342 + 0.1901$i$ \,&\, 1.334 + 0.1873$i$ \,&\, 1.326 + 0.1844$i$ \,&\, 1.318 + 0.1815$i$\, \\ \hline
 \, 3 \,&\, 1 \,&\, 1.321 + 0.5847$i$ \,&\, 1.314 + 0.5758$i$ \,&\, 1.307 + 0.5668$i$ \,&\, 1.299 + 0.5580$i$ \,&\, 1.292 + 0.5491$i$\, \\ \hline
 \, 3 \,&\, 2 \,&\, 1.270 + 0.9882$i$ \,&\, 1.264 + 0.9727$i$ \,&\, 1.259 + 0.9571$i$ \,&\, 1.253 + 0.9416$i$ \,&\, 1.247 + 0.9262$i$\, \\ \hline
 \, 3 \,&\, 3 \,&\, 1.204 + 1.402$i$ \,&\, 1.201 + 1.380$i$ \,&\, 1.197 + 1.357$i$ \,&\, 1.193 + 1.334$i$ \,&\, 1.189 + 1.312$i$ \,\\ \hline\hline
 \end{tabular}
 \caption{Quasinormal modes frequencies $(2M\omega)$ for the extreme potential case computed using third-order WKB approximation. The parameter ${{\scalebox{.9}{${\alpha}$}}}=0$ corresponds to the Schwarzschild case.}
 \label{tabela2}
\end{table}

\subsection{Quasinormal modes of GD hairy black holes: numerical results}
\label{numer}

The waveform resulting from the scalar perturbation of the decoupled hairy black hole, in the extreme case, will be numerically computed in this section \cite{Gundlach:1993tn,Gundlach:1993tp, Krivan:1996da}. It allows a precise account of the evolution of  quasinormal modes for a reliable range of the parameter ${{\scalebox{.9}{${\alpha}$}}}$. One can see that despite the significant deviation of the frequencies of the quasinormal modes for increasing ${{\scalebox{.9}{${\alpha}$}}}$, the wave deviation from the Schwarzschild case is relatively small, although it clearly indicates a signature of GD hairy solutions. 

\begin{figure}
\begin{subfigure}{0.49\textwidth}
    \includegraphics[width=\textwidth]{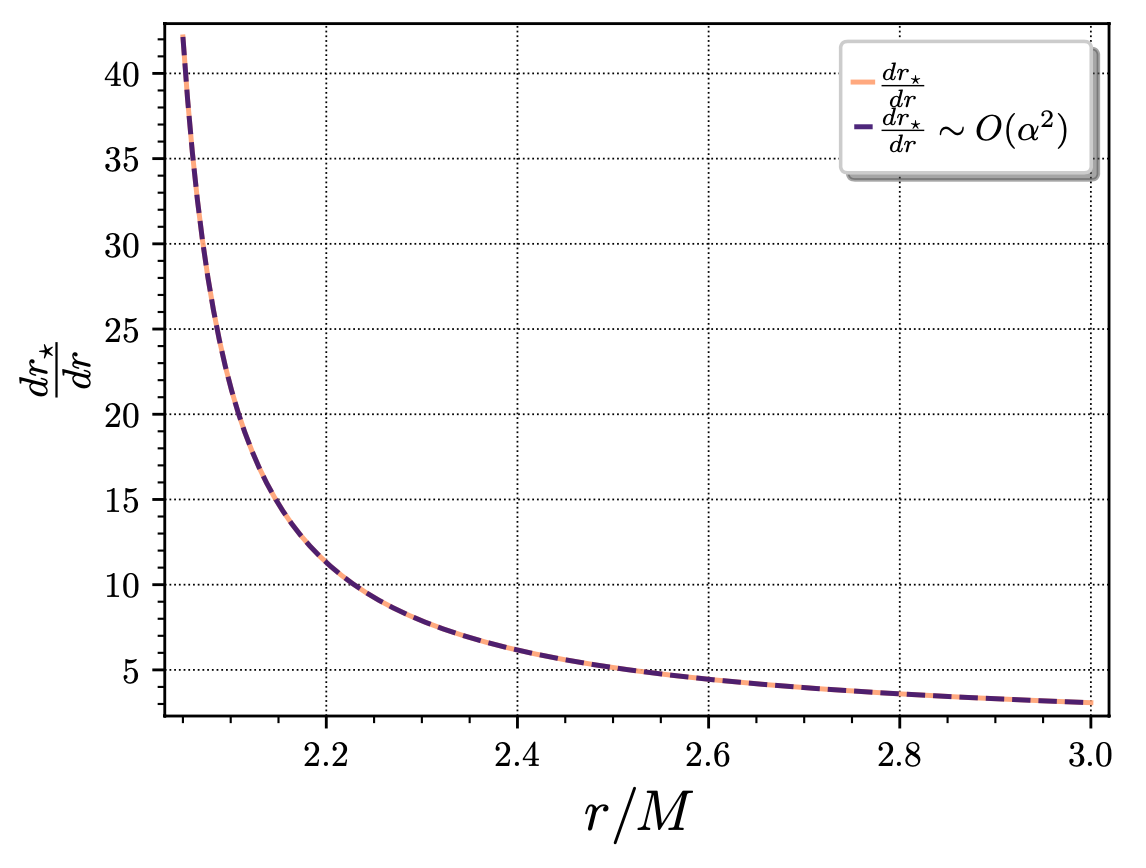}
 \caption{Behavior of $\dfrac{dr_\star}{dr}$ and its truncated series for ${{\scalebox{.9}{${\alpha}$}}}=0.2$.}
 \label{dtortoise1}
\end{subfigure}
\hfill
\begin{subfigure}{0.49\textwidth}
    \includegraphics[width=\textwidth]{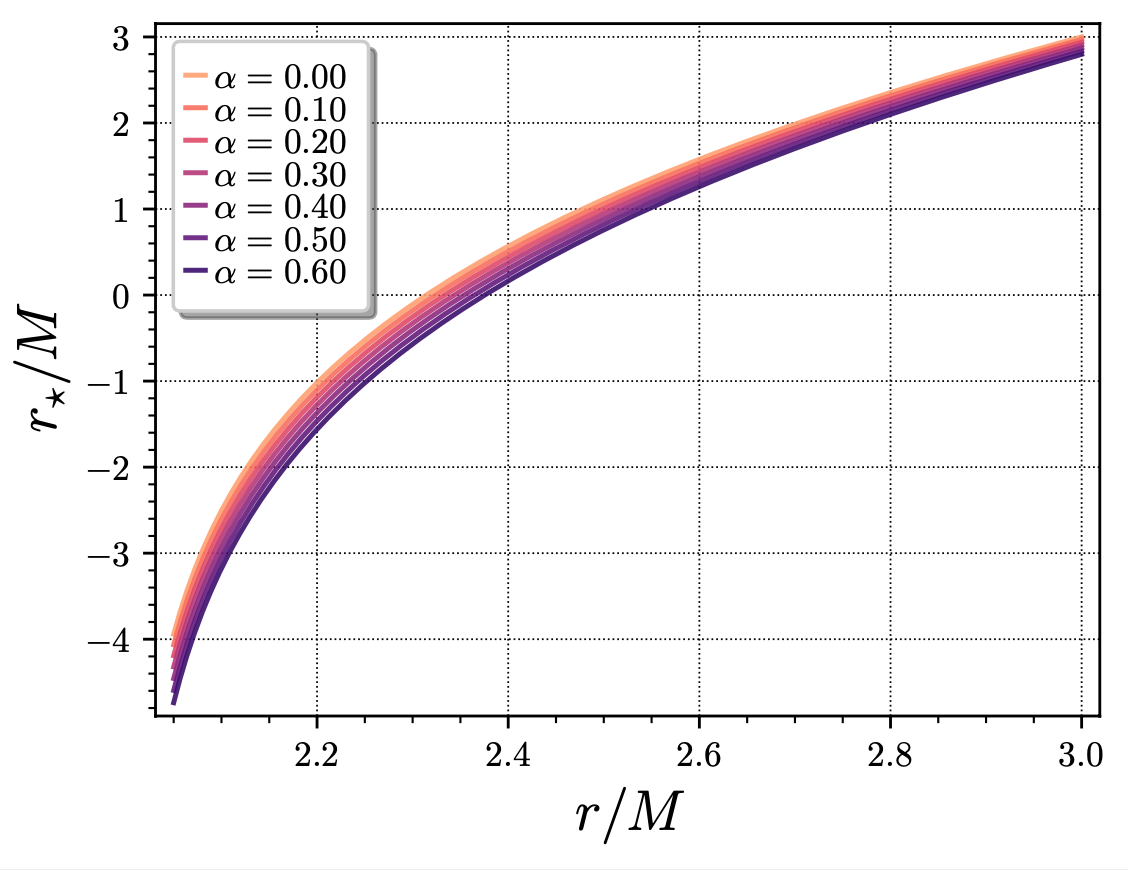}
 \caption{Plot of $r_\star$ with respect to $r$, for $0 \leq {{\scalebox{.9}{${\alpha}$}}} \leq 0.6$.}
 \label{dtortoise2}
\end{subfigure}
        
\caption{In both plots, the region main interest is the one close to the horizon, which is located at $2M$ in the extreme case.}
\label{fig:figures}
\end{figure}

Before solving the wave equation, as there is no analytical solution to Eq. (\ref{tortoise}), we expand its right-hand side up to the second order in ${{\scalebox{.9}{${\alpha}$}}}$, corresponding to the same order of the effective potential: 
\beq
\dfrac{dr_\star}{dr} &=& -\frac{r}{2M-r} + {{\scalebox{.9}{${\alpha}$}}} \frac{{\left(2Me^{-2}-re^{-\frac{r}{M}}\right)}r}{{\left(2M-r\right)}^{2}} -{{\scalebox{.9}{${\alpha}$}}}^{2} \frac{{\left(2Me^{-2}-re^{-\frac{r}{M}}\right)}^{2}r}{{\left(2M-r\right)}^{3}} + \mathcal{O}({{\scalebox{.9}{${\alpha}$}}}^3),
\eeq
which can be integrated analytically, up to $\mathcal{O}({{\scalebox{.9}{${\alpha}$}}}^3)$, yielding 
\begin{widetext}
\beq
r_\star &\approx & 2M \log\left(r -2 M\right) + r+{{\scalebox{.9}{${\alpha}$}}}\left[2 M e^{-2} \log\left(r -2 M\right) + \frac{4 M^{2} e^{-2}}{2 M - r} - \frac{4 M^{3} e^{-\frac{r}{M}} - M r^{2} e^{-\frac{r}{M}}}{4 M^{2} - 4 M r + r^{2}}\right]\nonumber\\\nonumber
&&
+{{\scalebox{.9}{${\alpha}$}}}^2 \left[4 M {\rm Ei}\left(2 - \frac{r}{M}\right) e^{-4} - 2 M {\rm Ei}\left(4 - \frac{2 r}{M}\right) e^{-4} + \frac{4 M^{2} e^{-4}}{2 M - r}\right. \\
&&
\left.- \frac{1}{2 {\left(4 M^{2} \!-\! 4 M r \!+\! r^{2}\right)}}\left(8 M^{3} e^{-4} \!+\! 16 M^{3} e^{-\frac{r-2}{M}} \!-\! 16 M^{2} r e^{-\frac{r-2}{M}} \!-\! 4 M^{3} e^{-\frac{2r}{M}} \!+\! 4 M^{2} r e^{-\frac{2r}{M}} \!+\! M r^{2} e^{-\frac{2r}{M}}\right)\right].\label{expansion1}
\eeq
\end{widetext}
Here $\mathrm{Ei}(z)$ denotes the exponential integral $\mathrm{Ei}(z) = \int_{-\infty}^z\frac{e^u}{u}du$. Truncating the $\dfrac{dr_\star}{dr}$ expansion up to $\mathcal{O}({{\scalebox{.9}{${\alpha}$}}}^3)$ does not affect the results, as the error in such an approximation is $\varepsilon \approx 0.00198\%$. Fig. \ref{dtortoise1} depicts these results, showing that assuming the truncated series for $\dfrac{dr_\star}{dr}$, in the expansion Eq. \eqref{expansion1}, is a completely reliable method for deriving the quasinormal modes of GD hairy black holes.  Fig. \ref{dtortoise2} shows that the GD parameter ${{\scalebox{.9}{${\alpha}$}}}$ forces the generalized tortoise coordinate to attaing higher values, for fixed values of the radial coordinate $r$. For all values of ${{\scalebox{.9}{${\alpha}$}}}$, the coordinate $r_\star$ is still a monotonically increasing function of $r$. However, the higher the value of ${{\scalebox{.9}{${\alpha}$}}}$, the steeper the coordinate $r_\star$ is, with respect to $r$.


\begin{figure}[H]
 \centering
\includegraphics[width=0.95\textwidth]{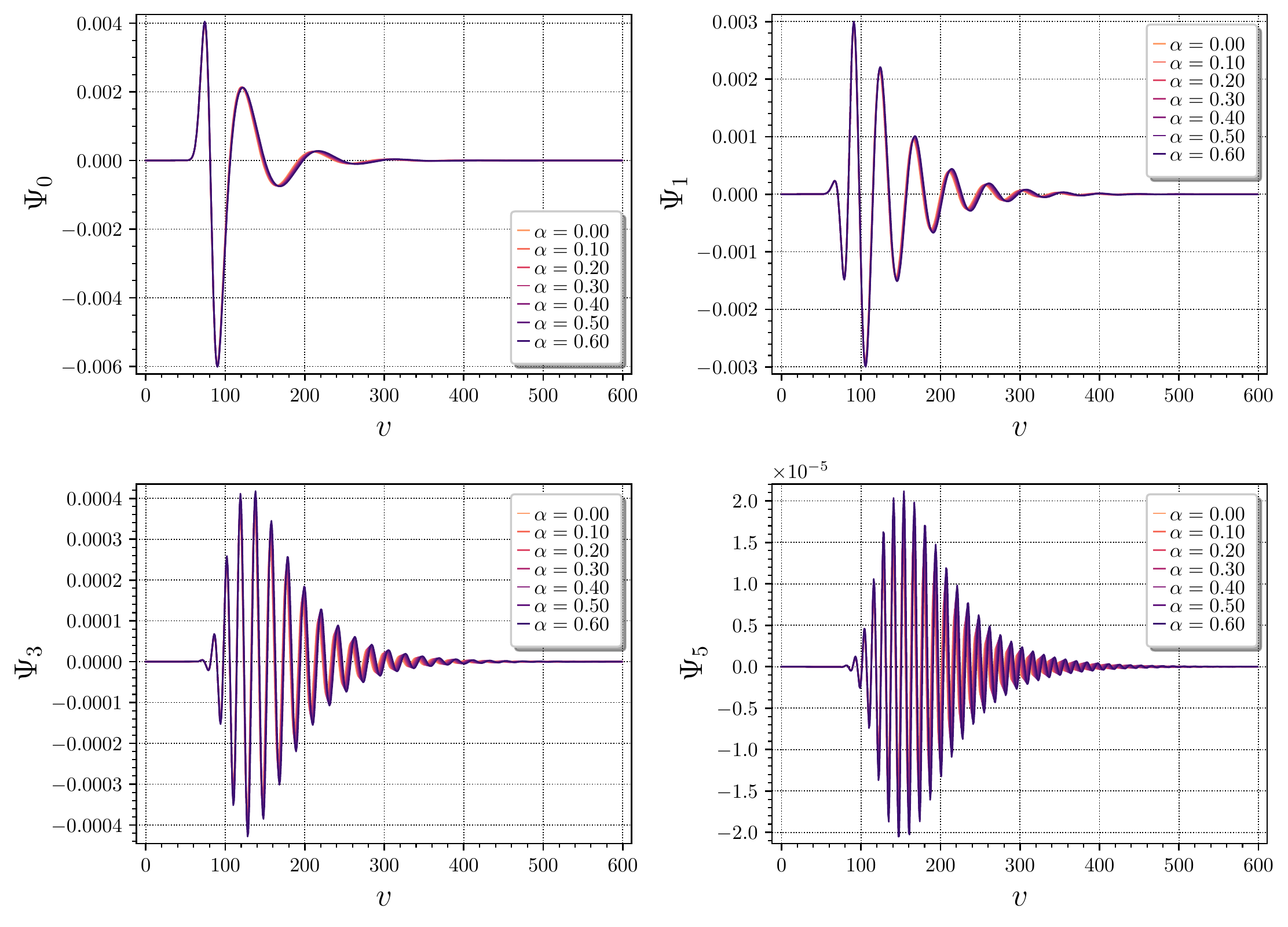}
 \caption{Wave profiles found as numerical solutions of Eq. (\ref{waveFun}), in light-cone coordinates, for $\ell = 0,1,3,5$ and several values of GD parameter ${{\scalebox{.9}{${\alpha}$}}}$.}
 \label{abswaves}
\end{figure}

Now the method can be properly applied. As introduced in Refs. \cite{Gundlach:1993tn,Gundlach:1993tp}, instead of apply the ansatz of Eq. (\ref{phiAnsatz}),  only the angular part of the Klein--Gordon equation is  separated. The resulting wavelike equation, for the extreme case, is given by
\begin{align}
 \frac{\partial^2\Psi_\ell}{\partial t^2} - \frac{\partial^2\Psi_\ell}{\partial r^2} +V_\ell(r,{{\scalebox{.9}{${\alpha}$}}})\Psi_\ell = 0,
\end{align}
where $V_\ell(r,{{\scalebox{.9}{${\alpha}$}}})$ is the same effective potential of Eq. (\ref{effPot2}). Such equation can be rewritten in terms of the light-cone coordinates 
\begin{align}
 du = dt - dr_\star, \qquad \qquad dv = dt + dr_\star,
\end{align}
yielding 
\begin{align}\label{waveFun}
 4\frac{\partial^2 \Psi_\ell}{\partial u \partial v} +V_\ell(r,{{\scalebox{.9}{${\alpha}$}}})\Psi_\ell= 0.
\end{align}
Eqs. \eqref{waveFun} can be discretized and solved numerically \cite{Gundlach:1993tn,Gundlach:1993tp, Konoplya:2011qq}. Expressing 
\begin{align}\label{eq11}
 \!\!\!\Psi(N) \!=\! \Psi(E)\!+\!\Psi(W)\!-\!\Psi(S)\!-\!\frac{h^2}{8} V_\ell\left(S\right)\left[\Psi(W)\!+\!\Psi(E)\right],
\end{align}
the points $N= (u+h,v+h)$, $S= (u,v)$, $E= (u,v+h)$ and $W= (u+h,v)$ form a null rectangle with relative position and $h$ is an overall grid scale. In order to apply Eq. (\ref{eq11}), a Gaussian wave will be employed as boundary condition, such that
\begin{subequations}
\begin{align}
 \Psi_\ell(v,-50) &= Ae^{-(v-B)^2/C^2}\\
 \Psi_\ell(-50,u) &= Ae^{-B^2/C^2}.
\end{align}
\end{subequations}
With $A=0.01$, $B=30$ and $C=3$. The results are shown in Figs. \ref{abswaves}, with wave profiles in a $10^3\times 10^3$ grid, for several values of $\ell$. The qualitative behavior resembles the Schwarzschild case, again represented by ${{\scalebox{.9}{${\alpha}$}}}=0$. However, as highlighted in subsequent plots, GD hairy black holes do have a unique signature, regarding the quasinormal modes, encoded in the GD parameter ${{\scalebox{.9}{${\alpha}$}}}$.


\begin{figure}[H]
 \centering
\includegraphics[width=0.95\textwidth]{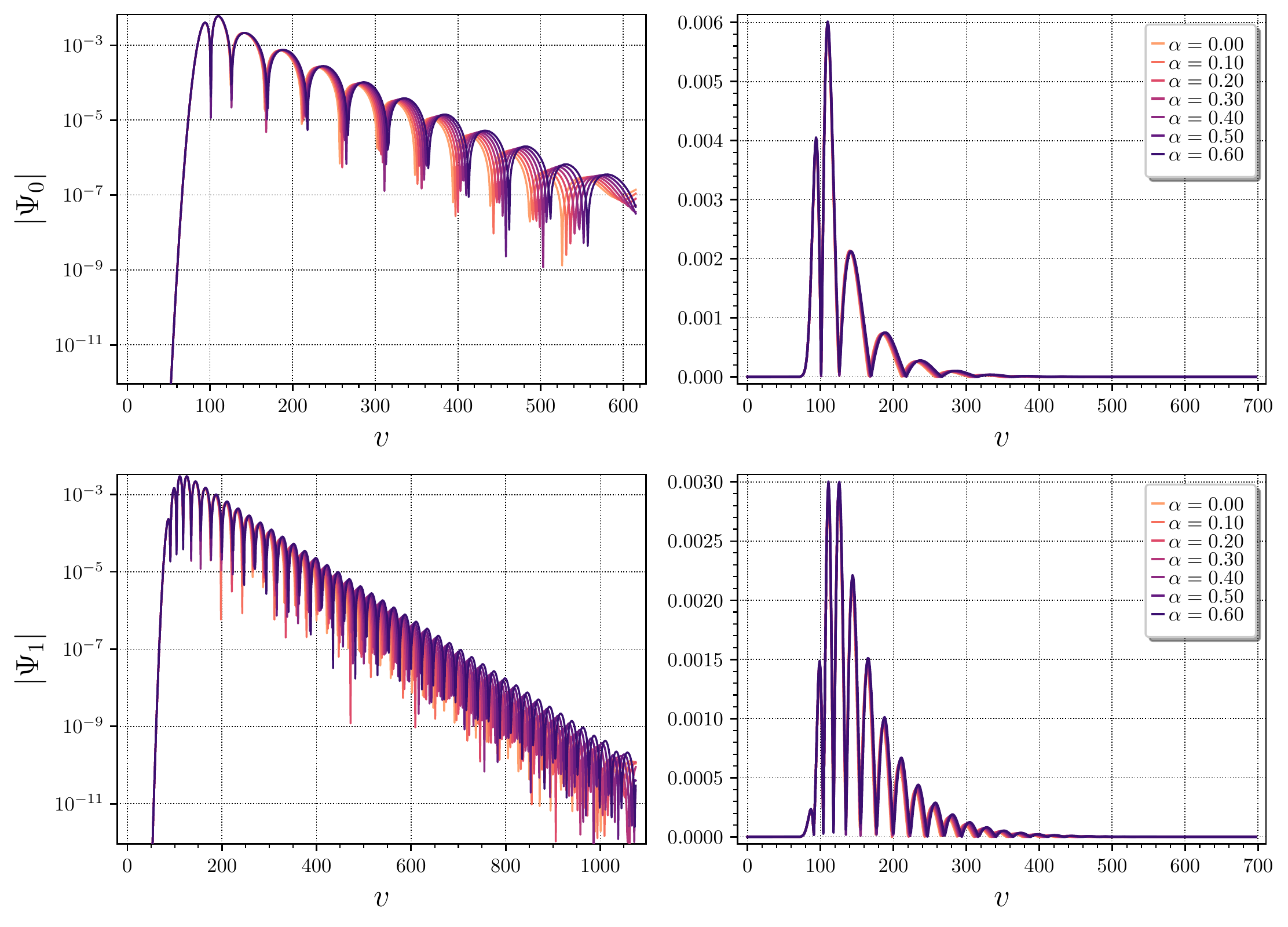}
 \caption{Absolute value of the wave profiles found as numerical solutions of Eq. (\ref{waveFun}), in light-cone coordinates, for a $10^3\times 10^3$ grid, $\ell=0,1$ and several values of GD parameter ${{\scalebox{.9}{${\alpha}$}}}$. The left panel depicts the wave profile in logarithm scale.}
 \label{abswaves1}
\end{figure}

Figs. \ref{abswaves} and \ref{abswaves1} show that the GD parameter ${{\scalebox{.9}{${\alpha}$}}}$ governing GD hairy black holes decreases the damping rate of the modes, irrespectively the value of $\ell$ is. It yields a GD hairy black hole that behaves as a better oscillator, when compared to the Schwarzschild standard case, corresponding to ${{\scalebox{.9}{${\alpha}$}}}=0$. Besides, for each fixed value of the GD parameter ${{\scalebox{.9}{${\alpha}$}}}$, the higher the value of the light-cone coordinate $v$, the faster the wave amplitude evanesces and the lower the amplitude is. It is a general behavior of the wave mode, also irrespectively of the value of $\ell$. Such results are in agreement with analytical solutions displayed in tables \ref{tabela} and \ref{tabela2}. As mentioned before, the imaginary part of the frequency $\omega _I$ is associated with the damping rate, which decreases for increasing ${{\scalebox{.9}{${\alpha}$}}}$. It has the side effect of amplifying $\Psi$ when compared to the Schwarzschild case. Also, again showing the agreement between the numerical and analytical results, the oscillation frequency decreases for increasing ${{\scalebox{.9}{${\alpha}$}}}$. It can be seen from tables \ref{tabela}, \ref{tabela2} and the absolute value of $\Psi_0$ in the logarithm scale, as displayed in the upper left panel of Fig. \ref{abswaves1}. Thus, despite the qualitative aspects of GD hairy black holes resembling the Schwarzschild ones, at least when regarding scalar perturbations, the qualitative results presented here show that there are categorical physical signatures that are intrinsic to GD hairy black holes and unequivocally different from the Schwarzschild standard ones.


\begin{figure}[H]
 \centering
\includegraphics[width=0.95\textwidth]{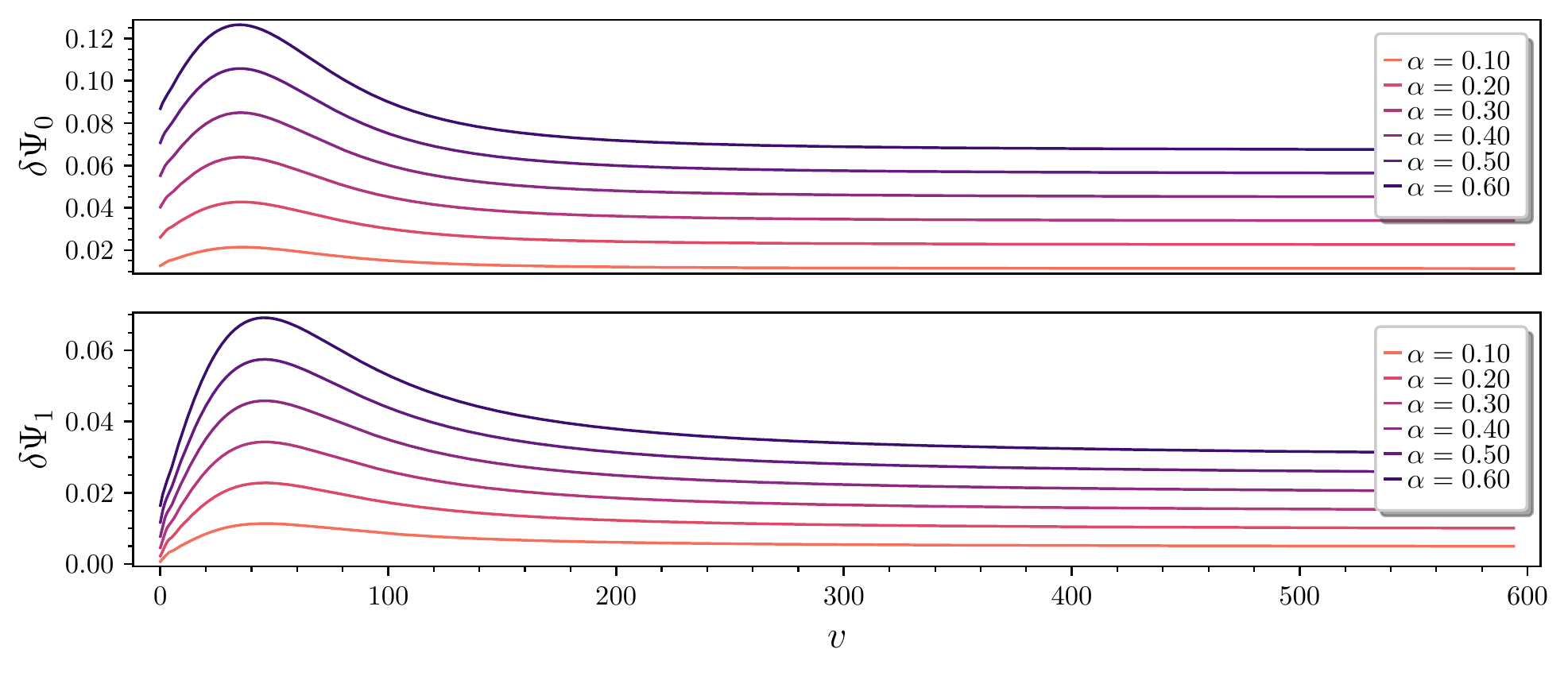}
 \caption{Relative difference of the wave profiles $\delta \Psi_\ell = \dfrac{|\Psi_\ell(\alpha)-\Psi_\ell(0)|}{|\Psi_\ell(0)|}$ for $\ell=0,1$ and several values of GD parameter ${{\scalebox{.9}{${\alpha}$}}}$. Here $\Psi_\ell(0)$ denotes to usual Schwarzschild case.}
 \label{abswaves3}
\end{figure}

Besides the points highlighted above, a clear signature of hairy black hole can also be seen from the relative difference of the wave profiles $\delta \Psi_\ell = \dfrac{|\Psi_\ell(\alpha)-\Psi_\ell(0)|}{|\Psi_\ell(0)|}$, as depicted in Fig. \ref{abswaves3}. There is a softened peak at the beginning of the signal, followed by a steady gap denoting the difference between the Schwarzschild and hairy black holes cases. Despite the very low amplitude of the wave profiles, the difference is substantial, strengthening our claim of a clear signature for scalar perturbations. We finish by emphasizing that the spectrum of quasinormal modes of GD hairy black holes represents the damping of the oscillations associated with the wave amplitude, being the fluctuation containing the lowest damping rate dominant at the late-time regime, whereas fluctuations containing a higher damping rate are usually suppressed according to an exponential law. In general, physical signatures of the quasinormal ringdown phase can be addressed when analyzing dominant modes with the lowest imaginary terms.    

\section{Conclusions and perspective}
\label{4}

The behavior of a scalar perturbation subjected to a hairy GD background solution was analyzed and compared to the results for a scalar field subjected to the ordinary Schwarzschild one. 
The quasinormal modes due to scalar perturbations of hairy GD black holes were derived and discussed, characterizing and dictating the late-time behavior of the scalar field. 
Initially, the Mashhoon method was applied to determine the quasinormal frequencies and their dependence on the additional parameters of the GD hairy black hole. The subsequent results were then calculated for the extreme case of the solution. The waveform and its characteristic damping behavior were derived numerically for a range of the hair parameter ${{\scalebox{.9}{${\alpha}$}}}$ and up to $\ell=5$. The peaks of the effective potential were also shown to scale with ${{\scalebox{.9}{${\alpha}$}}}$. The quasinormal modes for hairy GD black holes were shown to scale with the event horizon radius, at least for sufficiently large hairy GD black holes. As the scalar perturbation decay has a timescale that is inversely proportional to the eigenfrequency, it implies that the lower the hairy GD black hole mass, the more time elapses to approach an equilibrium end-stage. Also, quasinormal modes for hairy GD black holes were shown to scale with the event horizon radius, likewise the Bekenstein--Hawking law, stating that the temperature of a hairy GD black hole is proportional to its event horizon radius \cite{Cavalcanti:2022adb}. Thus the quasinormal modes eigenfrequencies vary with the temperature as well. Since Ref. \cite{Meert:2021khi} showed that GD hairy black holes can be accurately emulated in the membrane paradigm of AdS/CFT, one can still interpret the results here obtained in the light of Ref. \cite{Horowitz:1999jd}. In fact, large GD hairy black holes also correspond to thermal states in dual conformal 
field theory, whereas the scalar field decay matches the perturbation decay of the respective thermal state, yielding a reliable timescale for the behavior nearing thermal equilibrium.

\subsection*{Acknowledgements}

RCdP is grateful to the Coordenação de Aperfeiçoamento de Pessoal de Nível Superior -- Brasil (CAPES) for financial support, Finance Code 001.
RdR~is grateful to The S\~ao Paulo Research Foundation - FAPESP (Grants No. 2017/18897-8, No. 2021/01089-1, and No. 2022/01734-7) and the National Council for Scientific and Technological Development -- CNPq (Grants No. 303390/2019-0 and No. 402535/2021-9), for partial financial support.  
\bigskip

\textbf{Data Availability Statement:} No Data associated in the manuscript.

\bibliography{hairyGD_09_2022}

\end{document}